\documentclass[twocolumn]{aastex62}

\usepackage{color}
\newcommand{\source}{IGR\,J17591$-$2342}

\newcommand{\arcsecond}{$^{\prime\prime}$}

\newcommand{\nh}{$N_{\rm H}$}
\newcommand{\lrlx}{$L_{\rm R}$/$L_{\rm X}$}
\newcommand{\swift}{{\it Swift}}

\newcommand{\lr}{$L_{\rm R}$}
\newcommand{\lx}{$L_{\rm X}$}

\graphicspath{{./}{figures/}}

\received{2018}
\revised{2018}
\accepted{2018}
\submitjournal{ApJL}

\shorttitle{The radio-bright AMXP IGR~J17591$-$2342}
\shortauthors{T. D. Russell et al.}

\begin{document}

\title{The radio-bright accreting millisecond X-ray pulsar IGR J17591$-$2342}

\correspondingauthor{T. D. Russell}
\email{t.d.russell@uva.nl}

\author[0000-0002-7930-2276]{T. D. Russell}
\affil{Anton Pannekoek Institute for Astronomy, University of Amsterdam,   Science Park 904, 1098 XH, Amsterdam, The Netherlands}

\author{N. Degenaar}
\affil{Anton Pannekoek Institute for Astronomy, University of Amsterdam, Science Park 904, 1098 XH, Amsterdam, The Netherlands}

\author{R. Wijnands}
\affil{Anton Pannekoek Institute for Astronomy, University of Amsterdam, Science Park 904, 1098 XH, Amsterdam, The Netherlands}

\author{J. van den Eijnden}
\affil{Anton Pannekoek Institute for Astronomy, University of Amsterdam, Science Park 904, 1098 XH, Amsterdam, The Netherlands}

\author{N. V. Gusinskaia}
\affil{Anton Pannekoek Institute for Astronomy, University of Amsterdam, Science Park 904, 1098 XH, Amsterdam, The Netherlands}
\affil{ASTRON, Netherlands Institute for Radio Astronomy, Oude Hoogeveensedijk 4, 7991 PD, Dwingeloo, The Netherlands}

\author{J. W. T. Hessels}
\affil{Anton Pannekoek Institute for Astronomy, University of Amsterdam, Science Park 904, 1098 XH, Amsterdam, The Netherlands}
\affil{ASTRON, Netherlands Institute for Radio Astronomy, Oude Hoogeveensedijk 4, 7991 PD, Dwingeloo, The Netherlands}

\author{J. C. A. Miller-Jones}
\affil{International Centre for Radio Astronomy Research - Curtin University, GPO Box U1987, Perth, WA 6845, Australia}

\begin{abstract}
\source{} is a 527-Hz accreting millisecond X-ray pulsar that was discovered in outburst in 2018 August. In this paper, we present quasi-simultaneous radio and X-ray monitoring of this source during the early part of the outburst. \source{} is highly absorbed in X-rays, with an equivalent hydrogen absorption along the line of sight, \nh{}, of $\approx$4.4$\times$10$^{22}$\,cm$^{-2}$, where the Galactic column density is expected to be $\approx$1 -- 2$\times$10$^{22}$\,cm$^{-2}$. The high absorption suggests that the source is either relatively distant ($>$6\,kpc), or the X-ray emission is strongly absorbed by material local to the system. Radio emission detected by the Australia Telescope Compact Array shows that, for a given X-ray luminosity and for distances greater than 3\,kpc, this source was exceptionally radio loud when compared to other accreting neutron stars in outburst ($L_{\rm X} > 10^{33}$\,erg\,s$^{-1}$). For most reasonable distances, \source{} appeared as radio luminous as actively accreting, stellar-mass black hole X-ray binaries.

\end{abstract}


\section{Introduction}

Accreting millisecond X-ray pulsars (AMXPs) are binary systems in which a neutron star (NS) primary accretes material --- often only sporadically --- from a low-mass companion star. However, in contrast to the majority of known neutron star X-ray binaries (NSXRBs), these objects show X-ray pulsations as material is channelled from the disk onto the magnetic poles of the rotating NS \citep[e.g.][]{1998Natur.394..344W}. These objects are thought to be the link between accreting low-mass X-ray binaries (XRBs) and millisecond pulsars (MSPs; see \citealt{1998Natur.394..344W,2012arXiv1206.2727P}, for review) where the transfer of angular momentum during the accreting XRB phase can spin up the NS to millisecond periods (e.g. \citealt{1982Natur.300..728A,1982CSci...51.1096R}). This link has been exemplified, e.g., by the systems PSR J1023+0038 and
M28I (IGR J18245-2452), which have been observed to transition between
radio MSP and AMXP states on relatively short timescales \citep{2009Sci...324.1411A,2013Natur.501..517P,2014ApJ...790...39S}. Such transitional MSPs (tMSPs) are also thought to switch between accretion-powered and rotation-powered pulsar states.  

AMXPs are typically transient, going through cycles of outburst and quiescence (e.g. \citealt{1998A&A...331L..25I}). During outburst, radio emission with a flat or inverted spectrum ($\alpha \approx 0$, where the flux density, $S_{\rm \nu}$, scales with the frequency, $\nu$, such that  $S_{\rm \nu} \propto \nu^{\alpha}$) is observed \citep[e.g.][]{2017MNRAS.470..324T}, indicative of a partially self-absorbed, compact jet \citep[e.g.][]{1979ApJ...232...34B}.

In their hard X-ray states at the beginning and end of a typical outburst, black hole (BH) XRBs exhibit a tight, non-linear relationship between their radio ($L_{\rm R}$) and X-ray ($L_{\rm X}$) luminosities that extends over several decades in luminosity (e.g. \citealt[][]{2012MNRAS.423..590G,2013MNRAS.428.2500C}). The observed empirical relationship is generally described by two distinct power-law tracks in the radio/X-ray luminosity plane; an upper `radio-loud' track with a best fitting slope of $L_{\rm R}\propto L_{\rm X}^{\sim 0.6}$ and a lower, steeper, `radio-quiet' track with a best fitting slope of $L_{\rm R}\propto L_{\rm X}^{\sim 1}$ \citep{2012MNRAS.423..590G}. However, recently the statistical significance of the two separate tracks has been questioned \citep{2014MNRAS.445..290G,2018MNRAS.478L.132G}. Also, individual systems have shown significant deviation over luminosity ranges spanning $<$2 orders of magnitude \citep[e.g.][]{2013MNRAS.428.2500C}.

For accreting NSs the picture is even less clear. Not only is the connection complicated by the presence of a stellar surface/magnetosphere \citep[e.g.][]{2008ASPC..401..191M}, but also NSs appear fainter in the radio than their BH counterparts (e.g. \citealt{2001MNRAS.324..923F,2006MNRAS.366...79M}), meaning that their \lrlx{} behaviour has generally been measured across a smaller dynamic range (and with lower detection significance) than their BH counterparts. Initially, a steep $\sim$1.4 correlation was suggested \citep{2006MNRAS.366...79M}. However, individual systems appear to show varying correlation indices and normalisations, and not all  follow similar, or even well-defined tracks \citep[e.g.][]{2017MNRAS.470..324T}. To try to account for the restricted luminosity ranges typically measured for NS systems, regression analysis of a large sample of NS systems showed that they as a whole appear to be a factor of $\sim$22 fainter than the population of hard state BHs, but with a similar correlation slope \citep{2018MNRAS.478L.132G}.

The differences in both normalisation and slope for individual systems could be driven by a number of factors, such as jet power, primary mass, spin, magnetic field, and jet launching mechanism.  Recent results showing faint radio jets being launched from a slowly-spinning high-magnetic field ($>$10$^{12}$\,G) accreting NS indicate that the magnetic field or spin of the NS could be playing an important role in the radio brightness \citep{2018arXiv180910204V}.

\subsection{\source{}}
\source{} was discovered in outburst by the INTErnational Gamma-Ray Astrophysics Laboratory ({\it INTEGRAL}) on 2018 August 10--11 (MJD~58340--58341), during monitoring of the Galactic center \citep{2018ATel11941....1D}. Archival all-sky Neil Gehrels Swift Observatory (\swift) Burst Alert Telescope (BAT) observations indicate that the outburst rose above detection threshold around 2018 July 22 (MJD~58321), $\sim$20\,days earlier \citep{2018ATel11981....1K}. 
We observed the source at radio wavelengths using the Australia Telescope Compact Array (ATCA), detecting bright radio emission, leading to our initial suggestion that the source may be a BH candidate \citep{2018ATel11954....1R}. However, coherent X-ray pulsations at 527.4\,Hz (1.9\,ms) detected using the Nuclear Spectroscopic Telescope Array ({\it NuSTAR}) and the Neutron star Interior Composition Explorer ({\it NICER}) identified this source as an AMXP \citep{2018A&A...617L...8S}, with an orbital period of $\sim$8.8\,hours and a companion mass of $\gtrsim$0.42\,M$_\odot$. Assuming a 1.4\,M$_\odot$ NS mass and typical AMXP parameters, the spin frequency derivative implies a magnetic field of 1.4$\times$10$^8<$ B $<$8$\times$10$^9$\,G \citep{2018A&A...617L...8S}. 
After discovery, the source stayed active for $\sim$80\,days, exhibiting some long term variability, including multiple re-brightenings. 

In this paper, we discuss quasi-simultaneous radio and X-ray observations taken in the early stages of the outburst (spanning MJDs~58344--58356). The high X-ray absorption suggests that the source is relatively distant and therefore much more radio luminous than that from other known NS systems. We investigate the implications of this radio brightness in the context of other known XRBs.

\section{Observations}
\label{sec:observations}

\subsection{ATCA radio observations}
\label{sec:ATCA}

\begin{table*}
\caption{ATCA radio observations of \source{}. MJDs note the mid-point of the observation, with the quoted uncertainties reflecting the observation duration.}
\centering
\label{tab:ATCA}
\begin{tabular}{cccccc}
\hline
Date & MJD & Frequency & Flux density & $\alpha$\\
& & (GHz) & (mJy) & \\
\hline
2018 Aug 14 & 58344.56$\pm$0.12 & 5.5 & 1.09$\pm$0.02 & 0.1$\pm$0.3  \\ 
            &                   & 9.0 & 1.12$\pm$0.02 &  \\ 
2018 Aug 19 & 58349.26$\pm$0.07 & 5.5 & 1.00$\pm$0.05 & $-$0.30$\pm$0.25 \\ 
            &                   & 9.0 & 0.85$\pm$0.05 &    \\ 
2018 Aug 25 & 58355.28$\pm$0.12 & 5.5 & 1.18$\pm$0.04 & 0.25$\pm$0.1 \\ 
            &                   & 9.0 & 1.21$\pm$0.04 & \\ 
            & 58355.28$\pm$0.10 & 17.0 & 1.54$\pm$0.10 & \\ 
            &                   & 19.0 & 1.60$\pm$0.09 &    \\ 
\hline
\end{tabular}
\end{table*}

We conducted DDT radio observations (project ID: CX413) with the Australia Telescope Compact Array (ATCA). \source{} was observed three times early in its outburst, on 2018 August 14, 19, and 25 (MJDs 58344, 58349, and 58355, respectively; see Table~\ref{tab:ATCA}). The observations were taken at central frequencies of 5.5 and 9\,GHz. The August 25 observations had additional frequencies of 17 and 19\,GHz. Each frequency pair was taken simultaneously. All bands were recorded with a bandwidth of 2\,GHz. PKS 1934$-$638 was used for flux calibration, while phase calibration was carried out using 1817-254 on August 14 and with 1752-225 on August 19 and 25. The data were flagged and calibrated following standard procedures within the Common Astronomy Software Applications (CASA v4.7; \citealt{2007ASPC..376..127M}). 

For the August 14 observation the telescope was in a relatively extended (1.5D) configuration. Imaging was carried out with Briggs weighting (robust=0), minimising effects from diffuse emission in the field due to its location towards the Galactic center, providing angular resolutions of 6.8\arcsecond$\times$1.4\arcsecond at 5.5 GHz and 3.9\arcsecond$\times$0.9\arcsecond at 9 GHz (with a position angle 15\,degrees North of East). The flux density was determined by fitting for a point source in the image plane using IMFIT. We measure a (9\,GHz) right ascension (R.A.) and declination (Dec.) of:\\
\\
R.A. (J2000) =17${^{\rm h}}$59${^{\rm m}}$02${^{\rm s}}$.86 +/- 0.04, \\
Dec. (J2000) = -23${^{\circ}}$43${^{\rm m}}$08${^{\rm s}}$.3 +/- 0.1,\\
\\
where the uncertainties are from the fitted position of the point source, which are larger than the theoretical statistical error from beam centroiding. 

For our August 19 and 25 observations, ATCA was in a compact (H75) configuration, where five of the six antennas are located within 75\,meters of each other. Due to significant diffuse emission, imaging was not trivial. However, ATCA's sixth antenna is located 6\,km from the array core. Using only the longest baselines out to antenna six to eliminate issues from diffuse emission, we determined the flux by fitting for all sources in the field in the uv-plane using UVMULTIFIT \citep{2014A&A...563A.136M}.

See Table~\ref{tab:ATCA} for source flux densities.

\subsection{{\it Swift} X-ray observations}
The \swift{} X-ray Telescope (XRT) observed \source{} every few days during the outburst (target IDs: 10803 and 10804). Observations closest in time to our ATCA monitoring (within 1\,day) were extracted with the \swift-XRT online pipeline \citep{2009MNRAS.397.1177E}. We used the \swift-XRT online pipeline to create the lightcurve. The observations were conducted in photon counting (PC) mode. 

To determine the X-ray flux of \source{} for epochs that were closest to the radio observation, we analysed the 0.5--10\,keV spectra with \texttt{XSPEC} version 12.9 \citep{1996ASPC..101...17A}. The data were best fit with an absorbed \citep{2000ApJ...542..914W} powerlaw model; \texttt{Tbabs}$\times$\texttt{powerlaw} (C-stat/degrees of freedom = 269/240). The addition of a thermal component did not improve the fits, or change the results. X-ray fluxes were determined using the \texttt{XSPEC} convolution model \texttt{cflux}. Results are shown in Table~\ref{tab:XRT_flux}. The equivalent hydrogen column density, \nh, was tied between epochs (determined by jointly fitting all 17 \swift{} observations between MJDs~58342--58403). Leaving \nh{} free gave similar results for all but the August 18 observation, where the low count-rate meant poor model constraints. Fixing \nh{} to significantly different values produced poor fits and unphysical results.

\begin{table*}
\caption{Best-fitting spectral parameters for the 0.5--10\,keV \swift-XRT observations that were closest in time to the ATCA radio observations. The data were fit with \texttt{Tbabs}$\times$\texttt{powerlaw} within \texttt{XSPEC}. MJDs correspond to the mid-point of the X-ray observation (errors represent the observation duration). The column density along the line of sight, \nh, was tied between epochs to give \nh=(4.4$\pm$0.3)$\times$10$^{22}$\,cm$^{-2}$. $\Gamma$ is the photon index of the powerlaw component. Fluxes were calculated using \texttt{cflux}. Errors are 1-$\sigma$. 
}
\centering
\label{tab:XRT_flux}
\begin{tabular}{ccccccc}
\hline
& & & & Absorbed & Unabsorbed  \\
Date & MJD & Obs ID & $\Gamma$ & 1--10\,keV Flux & 1--10\,keV Flux  \\
&  & & & ($\times$10$^{-10}$)\,erg\,s$^{-1}$\,cm$^{-2}$ & ($\times$10$^{-10}$)\,erg\,s$^{-1}$\,cm$^{-2}$  \\
\hline
2018 Aug 14 & 58344.688$\pm$0.002 & 00010804002 & 1.55$\pm$0.15 & 1.98$^{+0.16}_{-0.15}$ &2.96$^{+0.17}_{-0.16}$  \\ 
2018 Aug 18 & 58348.223$\pm$0.003 & 00010804004 & 2.1$\pm$0.3 & 0.79$^{+0.11}_{-0.09}$ &1.43$^{+0.14}_{-0.13}$   \\ 
2018 Aug 25 & 58355.559$\pm$0.033 & 00010804006 & 2.0$\pm$0.15 & 1.84$\pm$0.09 &3.30$\pm$0.13  \\ 
\hline
\end{tabular}
\end{table*}

\section{Results and discussion}
\label{sec:results}

\subsection{Radio and X-ray lightcurves}

\begin{figure}
\centering
\includegraphics[width=0.9\columnwidth]{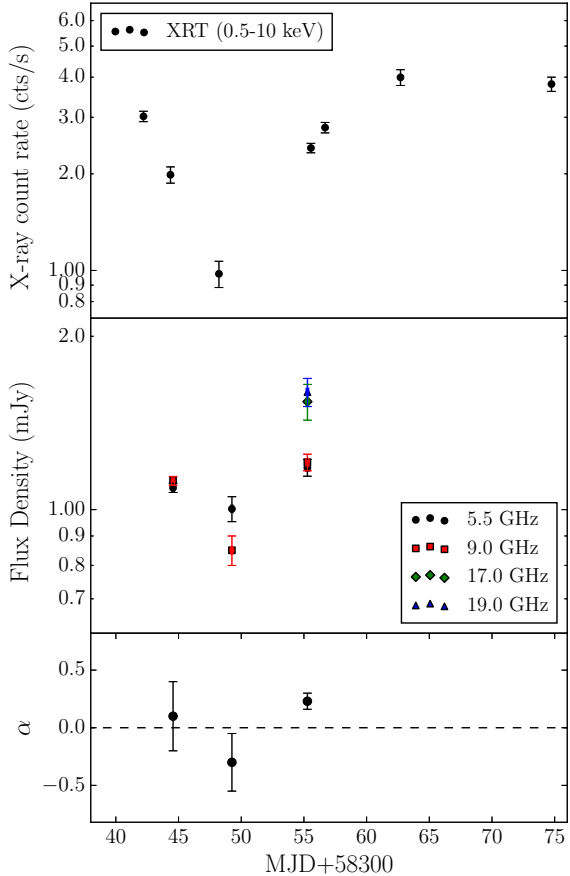}
\caption{X-ray and radio lightcurves of \source{} during the early stages of its 2018 outburst. {\it Top panel:} 0.5-10\,keV \swift-XRT observations. {\it Middle panel:} ATCA radio flux densities. {\it Bottom panel:} Radio spectral indices. Errors are 1-$\sigma$.}
\label{fig:lc}
\end{figure}

Following its initial X-ray detection, \source{} first brightened for a few days reaching an initial X-ray peak around August 14 (MJD~58344), around the time of the first \swift-XRT observation. \source{} then faded steadily over the next $\sim$5\,days (Figure~\ref{fig:lc}). The source then re-brightened, returning to a similar X-ray count rate as its initial X-ray peak, following which it remained detectable for the next $\sim$80\,days.

\begin{figure*}[!t]
\centering
\includegraphics[width=0.9\textwidth]{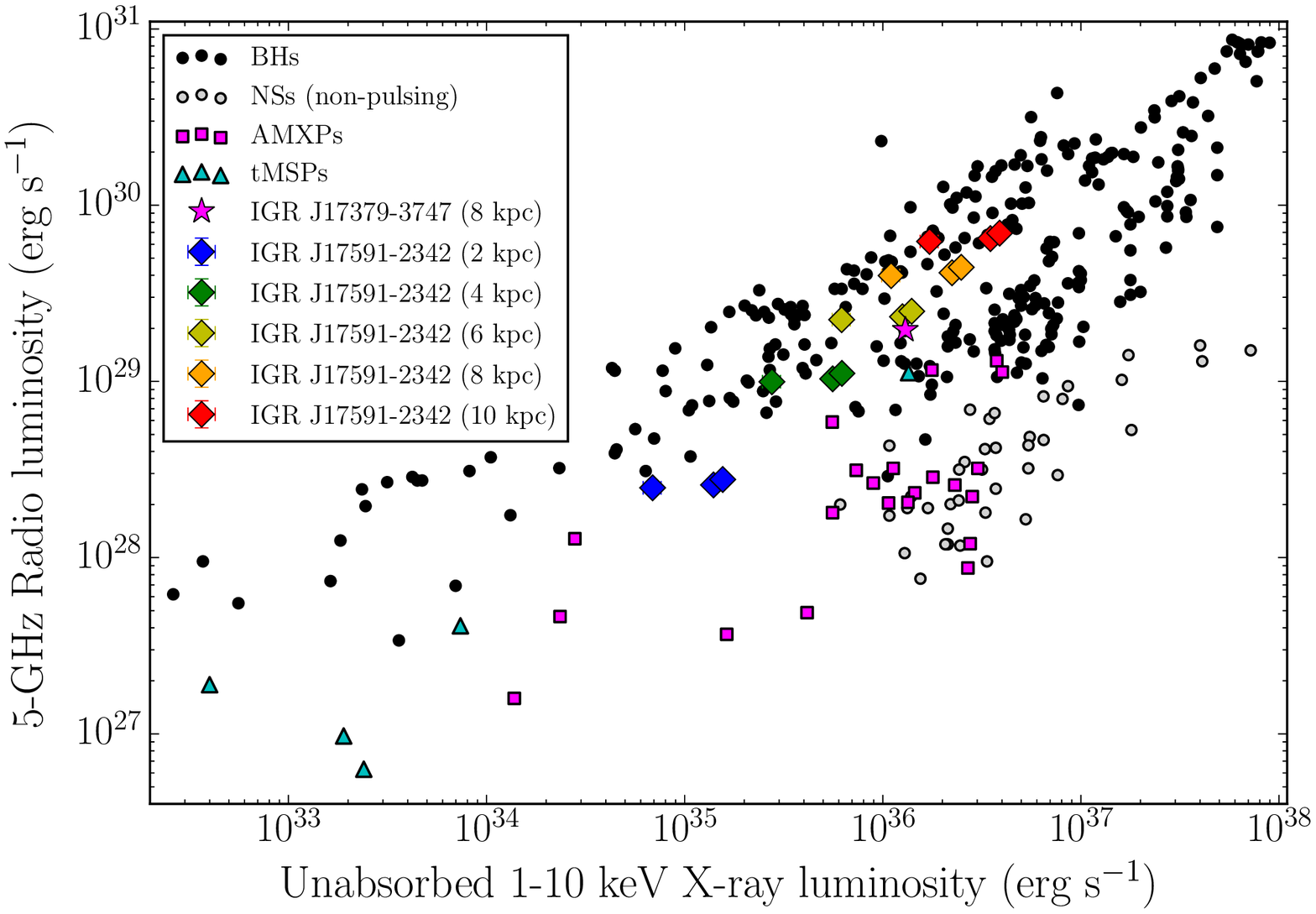}
\caption{Radio/X-ray measurements of \source{} taken during the early stages of its 2018 outburst. This figure is adapted from \citet{arash_bahramian_2018_1252036}, and shows the radio/X-ray observations for different types of accreting stellar-mass compact objects (often multiple observations for the same source). \source{} is shown by the diamonds, where colours show different assumed distances (see Section~\ref{sec:distances_lrlx}). We also show other AMXPs (magenta squares), highlighting IGR~J17379$-$3747 (magenta star) and tMSPs (cyan triangles). We see that at distances $\gtrsim$3\,kpc \source{} appeared more radio luminous than other NS systems at a similar X-ray luminosities. In fact, at such distances, \source{} lies in the region of parameter space usually occupied by black holes.}
\label{fig:lrlx_distances}
\end{figure*}

A few days after the initial X-ray peak, we conducted our first radio observation, detecting the source. Over the next $\sim$2\,weeks, we observed \source{} a further two times. During these observations, the source showed a similar evolution as the X-rays, where the radio flux faded briefly, before re-brightening (Figure~\ref{fig:lc}). 

During our first ATCA observation the radio spectrum was consistent with flat (Figure~\ref{fig:lc} and Table~\ref{tab:ATCA}). The spectrum steepened slightly at the time of our second radio observation, although signal to noise and frequency coverage meant poor constraints on the spectrum. During our final radio observation, we observed a significantly inverted radio spectrum. These observations imply partially self-absorbed synchrotron emission from a compact jet \citep[e.g.][]{2001MNRAS.322...31F}.

\subsection{Location on the radio/X-ray plane}
\label{sec:distances_lrlx}

To investigate the \lr{} and \lx{} of \source, we place the 5\,GHz radio and 1-10\,keV X-ray luminosities on the radio/X-ray plane using several assumed distances (Figure~\ref{fig:lrlx_distances}). Unfortunately, our contemporaneous observations sample only a very limited range of \lr{} and \lx{} (factors of only $\sim$1.1 and $\sim$2, respectively). Therefore, while we are unable to determine the specific scaling between the \lr{} and \lx{} with this monitoring, the radio brightness is interesting.

In terms of its radio and X-ray luminosities, for distances greater than $\sim$3\,kpc, \source{} would be brighter in the radio than the majority of other accreting NS systems (Figure~\ref{fig:lrlx_distances}). At a such distances, its \lrlx{} ratio places \source{} in the region of parameter space occupied by BHXRBs. These results imply that \source{} was either considerably radio-loud during the early stages of its 2018 outburst, or it is closer than 3\,kpc. 

Comparing the observed radio emission from \source{} to the sample of AMXPs with reported radio detections (Figure~\ref{fig:lrlx_distances}, magenta squares), we see that \source{} appears to be more radio bright than all other NS systems for distances greater than $\sim$5\,kpc, where this reduces to $\sim$3\,kpc without the inclusion of the AMXP IGR~J17379$-$3747. The recent radio detection of IGR~J17379$-$3747 \citep{2018ATel11487....1V} indicate this as another potentially radio bright AMXP (Figure~\ref{fig:lrlx_distances}, magenta star), where an 8\,kpc distance upper-limit was estimated from type I X-ray bursts, assuming the peak flux of the bursts reached the empirical NS Eddington limit.

\subsection{Absorption along the line of sight}
\label{sec:distances_nh}

The soft X-ray emission observed from \source{} is highly absorbed. Using ISM abundances from \citet{2000ApJ...542..914W}, we find a best fit from \swift-XRT data of \nh=(4.4$\pm$0.3)$\times$10$^{22}$\,cm$^{-2}$. This is consistent with results presented by \citet{2018A&A...617L...8S}, where the \nh{} was determined to be (3.6$\pm$1.1)$\times$10$^{22}$\,cm$^{-2}$ from {\it INTEGRAL}, {\it  NuSTAR}, and \swift-XRT observations taken at similar times. Our result is marginally higher than that found using the {\it NICER} data taken on similar dates (August 15 and August 18), which gave (3.45$^{+0.18}_{-0.15}$)$\times$10$^{22}$\,cm$^{-2}$ and (3.59$^{+0.05}_{-0.08}$)$\times$10$^{22}$\,cm$^{-2}$, respectively \citep{2018A&A...617L...8S}, agreeing at the $\sim$2-$\sigma$ level. Chandra observations taken on August 23 find \nh{}=(4.9$\pm$0.2)$\times$10$^{22}$\,cm$^{-2}$ \citep{2018ATel11988....1N}, consistent within the errors of our measured value.

\source{} lies in the direction of the Galactic center \citep{2018ATel11941....1D}, with an unknown distance to the source. However, our modelled \nh, which traces the interstellar gas along the line of sight, is $\sim$3--4 times higher than the expected Galactic contribution of \nh$_{\rm Gal}$=1.12--1.44$\times$10$^{22}$\,cm$^{-2}$ \citep{1990ARA&A..28..215D,2005A&A...440..775K,2013MNRAS.431..394W}. While a difference between the measured and expected \nh{} does not indicate the distance to the source, and the results may be dependent on the spectral model, it suggests that the source is not nearby. Converting between \nh{} and optical extinction \citep{2015MNRAS.452.3475B}, we can also compare against higher resolution three-dimensional optical reddening maps \citep{2018MNRAS.478..651G}. From these maps we also find that the measured absorption is significantly higher than the expected line of sight Galactic reddening, where the total Galactic contribution from reddening maps corresponds to $\sim$2.2$\times$10$^{22}$\,cm$^{-2}$. In fact, along this line of sight, for a distance of $<$6\,kpc, the Galactic contribution is expected to only be $\lesssim$7$\times$10$^{21}$\,cm$^{-2}$.

To further test the relationship with source distance, we compared the \nh$_{\rm Gal}$ \citep{1990ARA&A..28..215D,2005A&A...440..775K,2013MNRAS.431..394W} to the published\footnote{To be consistent, when possible, \nh{} from the literature were determined with \texttt{Tbabs}.} \nh{} for the 28 BH and NS XRBs with well estimated source distances\footnote{To test for differences between nearby and more distant objects.} \citep[][and references therein]{2018arXiv180411349G,2014PASA...31...16M}. We find that in all cases but two (GRS~1915+105 and V4641~Sgr) the measured \nh{} was either below or comparable to the expected Galactic line-of-sight absorption for these systems.

GRS~1915$+$105 and V4641~Sgr are both atypical and relatively distant BHXRBs ($>$6\,kpc), where the measured X-ray absorption exceeds the expected line of sight absorption by a factor of $\sim$4--8 due to anomalously high elemental abundances \citep{2001ApJ...555..489O,2002ApJ...567.1102L}. For these systems, the overabundances are thought to be related to material in or near the local environment of the source \citep[e.g.][]{1996ApJ...472L.111M,2000A&A...356..943M,2002ApJ...567.1102L}. Other systems with unknown distances also exhibit higher than expected absorption, however, this is generally ascribed to their close proximity to the Galactic center (estimated from the high absorption; e.g. \citealt{2006ApJ...646..394M}), or are high-inclination `dipping' sources \citep[e.g.][]{2002A&A...386..910P}, where disk material or the donor star passes through the line of sight. However, \source{} does not display X-ray dips \citep{2018A&A...617L...8S}, which are observed from high inclination systems, implying inclination is not the cause for the high \nh.

There is some debate over the origin of the dominant X-ray absorber for XRBs, whether it be the ISM (even with highly ionized winds from the XRB; \citep[e.g.][]{2009ApJ...707L..77M}, or from gas intrinsic to the XRB \citep[e.g.][]{2014ApJ...780..170L}. However, from the absorption, we favour the scenario in which \source{} is not nearby ($>$6\,kpc), likely close to the Galactic center (where the high absorption may arise from within the Galactic bulge). At such a distance \source{} is very radio-loud, with an \lr{} comparable to those observed from BHXRBs. This result identifies limitations when classifying XRBs from their \lrlx{} ratio. It is important to note that the a some of the \nh{} may arise from absorbing material local to the system, similar to what has been proposed for GRS~1915+105 and V4641~Sgr. While this would allow \source{} to be more nearby, we do not expect it closer than $\sim$3\,kpc, making this a radio-loud NS. High-resolution X-ray spectroscopy is needed to identify element overabundances from absorption local to the source \citep{2018ATel11988....1N}.

\subsection{Why is \source{} radio bright?}

While flaring from a transient jet or possibly a propeller mechanism could enhance the observed radio emission \citep[e.g.][]{2017MNRAS.470..324T}, our observations suggest that the radio emission originated from a steady, compact jet. 

Our first and third radio epochs showed a flat-to-inverted radio spectrum, inconsistent with emission from a transient jet (which would produce a steep spectrum). While the radio spectrum may have been steep during our second epoch, the flux remained stable over the three epochs arguing against significant flaring. Additionally, during the first epoch\footnote{Which was the only epoch we could test (Section~\ref{sec:ATCA}).} we did not detect significant intra-observation variability.

Furthermore, using system parameters determined by \citet{2018A&A...617L...8S} and typical AMXP values, we expect a magnetic propeller regime to occur at an \lx$\lesssim$8$\times$10$^{34}$\,erg\,s$^{-1}$ \citep{2017A&A...608A..17T}. Therefore, \source{} would need to be at distances $\lesssim$1.5\,kpc for the observed radio emission to be propeller driven, which we find unlikely (Section~\ref{sec:distances_nh}).

There is thought to be some dependence between jet emission and spin for an accreting NS \citep{2016ApJ...822...33P}. \source{} is one of the more rapidly spinning AMXPs (527.4\,Hz; \citealt{2018A&A...617L...8S}). However, comparing its spin and radio luminosity with other AMXPs, there does not appear to be a clear relationship between these two properties. Therefore, while the NS spin may play some part in enhancing the jet emission in \source, other characteristics, such as the magnetic field, NS mass, inclination, jet velocity, or launching mechanism, must play an important role in the observed jet emission of other systems (although the magnetic field of \source{} is typical of AMXP systems; \citealt{2018A&A...617L...8S}).

\section{Conclusions}

Using quasi-simultaneous observations, we explore the \lrlx{} of \source{} during the early stages of its 2018 outburst. We find that the source is highly absorbed in the soft X-ray band. Therefore, either \source{} is relatively distant and exceptionally radio-loud, or the source is more nearby with a significant contribution to the absorption from material local to the source, similar to the BHXRBs GRS~1915+105 and V4641~Sgr. If \source{} is at $>$3\,kpc, then our observations place it at radio luminosities more similar to those observed from BHXRBs than NS systems, which could mean that such radio-loud NSs would be a source of confusion when classifying XRBs from their \lrlx.

\acknowledgments
We thank Jamie Stevens and ATNF for rapid scheduling of the ATCA observations. TDR thanks Tobias Beuchert, Alice Borghese, Guglielmo Mastroserio, Pikky Atri, Juan Hern\'andez Santisteban and the Jetset group for helpful discussions. TDR acknowledges support from the Netherlands Organisation for Scientific Research (NWO) Veni Fellowship. ND and JvdE are supported by a NWO Vidi grant, awarded to ND. NVG acknowledges funding from NOVA. JWTH acknowledges funding from an NWO Vidi fellowship and from the European Research Council under the European Union’s Seventh Framework Programme (FP/2007-2013)/ERC Starting Grant agreement no. 337062 (‘DRAGNET’). JCAM-J is the recipient of an Australian Research Council Future Fellowship (FT140101082). The International Centre for Radio Astronomy Research is a joint venture between Curtin University and the University of Western Australia, funded by the state government of Western Australia and the joint venture partners. The Australia Telescope Compact Array is part of the Australia Telescope National Facility which is funded by the Australian Government for operation as a National Facility managed by CSIRO. We acknowledge the use of public data from the Swift data archive. This research has made use of NASA's Astrophysics Data System (ADS).

\vspace{5mm}
\facilities{ATCA, \swift-XRT}

\software{CASA, XSPEC}

\end{document}